\documentclass[epj,final]{svjour}
\usepackage{graphicx,amssymb,amsmath}

\newif\ifpsfrag
\psfragtrue
\ifpsfrag
  \usepackage{psfrag}
\fi

\def\ket#1{|{#1}\rangle}
\def\bra#1{\langle{#1}|}

\def\mysmall{\fontsize{6.5}{6.5}}

\begin{document}
\def\be{\begin{equation}}
\def\ee{\end{equation}}
\def\ba{\begin{eqnarray}}
\def\ea{\end{eqnarray}}

\title{Numerical evidence of the double-Griffiths phase
of the random quantum Ashkin-Teller chain}

\author{C. Chatelain \and D. Voliotis}
\institute{
Groupe de Physique Statistique,
D\'epartement P2M,
Institut Jean Lamour (CNRS UMR 7198),
Universit\'e de Lorraine, France
\email{christophe.chatelain@univ-lorraine.fr}}

\authorrunning{C. Chatelain \and D. Voliotis}
\titlerunning{Griffiths phases of the random Ashkin-Teller chain}

\abstract{
  The random quantum Ashkin-Teller chain is studied numerically by
  means of time-dependent Density-Matrix Renormalization Group.
  The critical lines are estimated as the location of the peaks of
  the integrated autocorrelation times, computed from spin-spin
  and polarization-polarization autocorrelation functions. 
  Disorder fluctuations of magnetization and polarization are
  observed to be maximum on these critical lines. Entanglement entropy
  leads to the same phase diagram, though with larger finite-size effects.
  The decay of spin-spin and
  polarization-polarization autocorrelation functions provides
  numerical evidence of the existence of a double Griffiths phase
  when taking into account finite-size effects. The two associated
  dynamical exponents $z$ increase rapidly as the critical lines
  are approached, in agreement with the recent conjecture of a
  divergence at the two transitions in the thermodynamic limit.
  }

\maketitle

\section{Introduction}
\label{Sec1}
Classical and quantum phase transitions are affected differently by
the introduction of homogeneous disorder. In the former, it is well
established that, when no frustration is induced, disorder is a relevant
perturbation at a critical point when thermal fluctuations grow slower
than disorder ones inside the correlation volume. It follows that the
critical behavior is unchanged when the specific heat exponent
$\alpha$ of the pure model is negative~\cite{Harris}. This criterion,
due to Harris, has been extensively tested on classical toy models such
as the 2D Ashkin-Teller model~\cite{Domany} or the 2D $q$-state Potts
model~\cite{RBPM,RBPM2}. In the latter, disorder is relevant for $q>2$
and the new random fixed point depends on the number of states $q$.
\\

Quantum phase transitions, i.e. transitions driven by quantum fluctuations
rather than thermal ones, involve new phenomena. First, randomness
can never be considered as homogeneous because time plays the role of an
additional dimension. Therefore, in contrast to the classical case,
even when random couplings are homogeneously distributed on the lattice,
they are always infinitely correlated in the time direction. Indeed,
the random quantum Ising chain in a transverse field (RTFIM),
for instance, is equivalent to the celebrated McCoy-Wu model,
a classical 2D Ising model with couplings that are randomly distributed
in one direction but perfectly correlated in the second one~\cite{McCoy,McCoy2}.
As a consequence, scale invariance is broken even after averaging over
disorder. The random quantum fixed point is usually invariant under
anisotropic scaling transformations. Correlation length $\xi$ and
autocorrelation time $\xi_t$ grow differently when approaching
the random quantum critical point:
     \begin{equation}
     \xi_t\sim \xi^z,
     \label{Defz}
     \end{equation}
where $z$ is the dynamical exponent. In the RTFIM, or in any model whose
critical behavior is described by the same fixed point, this dynamical
exponent increases algebraically when approaching the critical point
and diverges at the critical point.
\\

Another feature of quantum phase transitions in presence of disorder
is the existence of Griffiths phases~\cite{Griffiths}.
In the paramagnetic phase,
there may exist large regions with a high concentration of strong
couplings which can therefore order ferromagnetically earlier than
the rest of the system. Even though the probability of such regions
is exponentially small, they can cause a singular behavior of the free
energy with respect to the magnetic field in a finite range of values
of the quantum control parameter. The region of the paramagnetic phase
where this phenomena occurs is called a disordered Griffiths phase.
A similar phenomena takes place in the ferromagnetic phase. The
singular behavior is due to regions of the system with a high
concentration of weak bonds at their boundaries. They are therefore
only weakly coupled to the rest of the system and can order
independently~\cite{VojtaReview}. Because the tunneling time of these
rare regions grows exponentially fast with their size, they have a drastic
effect on the average autocorrelation functions of the system. Instead
of the usual exponential decay, the latter displays an algebraic
decay~\cite{RiegerYoung}
  \begin{equation}
    \overline{A(t)}\sim t^{-1/z}
    \label{AutocorrGrf}
  \end{equation}
involving the dynamical exponent $z$. In classical systems, Griffiths
phases usually consist in essential singularities, too weak to be observed
numerically, apart with some long-range correlated
disorder~\cite{ChatelainAT,ChatelainATb}.
\\

The quantum Ising chain in a transverse magnetic field has been, by far,
the most studied system undergoing a quantum phase transition. The
mapping of this model onto a lattice gas of free fermions allowed for exact
calculations in the pure case~\cite{Lieb}. In the presence of random
couplings, exact results are sparse~\cite{Shankar} but the mapping still allows
for an efficient numerical estimate of static, as well as dynamic,
quantum averages~\cite{Young}. The critical behavior is governed by
an unusual infinite-randomness fixed point (IRFP) which has been
extensively studied using a real-space renormalization group approach,
the Strong-Disorder Renormalization Group (SDRG), first introduced
by Ma and Dasgupta~\cite{MaDasgupta}, and later extended to the
RTFIM by Fisher~\cite{Fisher,Fisher2,Monthus}.
The strongest coupling, exchange interaction or transverse field, is
decimated by projecting out the Hilbert space onto the ground state of
this coupling. Other couplings are then treated using second-order
perturbation theory. Nevertheless, the method is believed to become
exact as the IRFP is approached because
the probability distribution of random couplings becomes broader and
broader and therefore, a strong coupling is always surrounded by weaker
couplings that can be treated perturbatively. The dynamical exponent
$z$ was shown to diverge at the phase transition.
The relation (\ref{Defz}) is replaced by $\xi\sim (\ln\xi_t)^{1/\psi}$
with $\psi=1/2$. Autocorrelation functions decay as~\cite{Igloi}
   \begin{equation}
     \overline{A(t)}\sim (\ln t)^{-2x_\sigma}
     \label{AutocorrFP}
   \end{equation}
at the critical point,
while correlation functions $C(r)$ display a more usual algebraic decay
with the distance $r$. The Ma-Dasgupta renormalization group allows for
the exact determination of the magnetization scaling dimension
and the correlation length exponent~\cite{Fisher,Fisher2}:
   \begin{equation}
     2x_\sigma={2\beta/\nu}=2-{1+\sqrt 5\over 2},
     \quad \nu=1/\psi=2.
     \label{ExposantIRFP}
   \end{equation}
The approach has been applied numerically to higher dimensions~\cite{Kovacs,Kovacs2,Kovacs3}.
The IRFP of the RTFIM is quite robust: in contrast to the classical case,
the random quantum $q$-state Potts chain falls also into this universality
class for any value of $q$~\cite{Senthil,CarlonPotts}.
\\

In this paper, a model with a richer phase diagram is considered.
The quantum two-color Ashkin-Teller model can be seen as two coupled Ising
chains in a transverse field. The Hamiltonian is~\cite{Kohmoto}
   \begin{eqnarray}
     H=&&-\sum_i \big[J_i \sigma_i^z\sigma_{i+1}^z+h_i\sigma_i^x\big]
     -\sum_i \big[J_i \tau_i^z\tau_{i+1}^z+h_i\tau_i^x\big]
     \nonumber\\
     &&\quad\quad\quad
     -\sum_i \big[K_i \sigma_i^z\sigma_{i+1}^z\tau_i^z\tau_{i+1}^z
     +g_i\sigma_i^x\tau_i^x\big]
     \end{eqnarray}
where $\sigma_i^{x,y,z}$ and $\tau_i^{x,y,z}$ are two sets of
Pauli matrices. The model possesses two ${\mathbb Z}_2$-symmetries,
corresponding to the invariance of the Hamiltonian under the reversal
of all spins $\sigma_i$ (or $\tau_i$) and of both $\sigma_i$ and
$\tau_i$. The breaking of these symmetries can be monitored using the
two order parameters
   \begin{equation}
    M=\sum_i\langle\sigma_i^z\rangle,\quad
    P=\sum_i\langle\sigma_i^z\tau_i^z\rangle
    \end{equation}
referred to as magnetization and polarization. In the pure case, i.e.
$J_i=J$, $K_i=K$, $h_i=h$ and $g_i=g$, the phase diagram involves several
critical lines, as the 2D classical Ashkin-Teller model~\cite{AshkinTeller,%
Fan,Fan2,Kamieniarz}.
When $K<J$, the two ${\mathbb Z}_2$ symmetries are simultaneously broken
and the Ashkin-Teller model undergoes a single second-order quantum phase
transition with the control parameter $\delta=J/h$. The scaling dimensions
of magnetization, polarization and energy densities vary along the
critical line~\cite{Kohmoto}:
     \begin{equation}
       x_\sigma={1\over 8},\ 
       x_{\sigma\tau}={\pi\over 8\arccos(-\epsilon)},\ 
       x_\varepsilon={\pi\over 2\arccos(-\epsilon)}
     \end{equation}
for $\epsilon=K/J\in[-1/\sqrt 2;1]$. For $K>J$, i.e. $\epsilon>1$, the
critical line splits into two lines, both belonging to the Ising universality
class ($x_\sigma=1/8$, $x_{\sigma\tau}=x_{\sigma^2}=1/16$ and $x_\varepsilon=1$).
These lines separate the paramagnetic ($M=P=0$) and Baxter ($M,P\ne 0$) phases
from an intermediate mixed phase ($M=0$, $P\ne 0$).
\\

In the following, the random Ashkin-Teller chain is considered. The four
couplings $J_i$, $h_i$, $K_i$ and $g_i$ are random variables, though not
independent but constrained by the relation~\footnote{
  The case where $K_i/J_i=\epsilon_J$ and $g_i/h_i
  =\epsilon_h$ are different was considered in~\cite{Vojta1}.
  At the infinite-randomness fixed point, both quantities are
  renormalized to the same value, $\epsilon^*=0$ in the weak-coupling
  regime ($\epsilon_J,\epsilon_h<1$) and $\epsilon\rightarrow +\infty$
  in the strong-coupling one. Without loss of generality, one can
  start with $\epsilon_J=\epsilon_h$. The more general case where
  $\epsilon_J$ and $\epsilon_h$ are random variables and are allowed
  to take values both above and below 1 was also considered in~\cite{Vojta1}
  and leads to a different critical behavior at the
  multicritical point.
}
    \begin{equation}
      {K_i\over J_i}={g_i\over h_i}=\epsilon
      \label{DefEpsilon}
    \end{equation}
where $\epsilon$ is a site-independent fixed parameter. This model was first
studied numerically by means of Density-Matrix Renormalization Group (DMRG)
in the weak-disorder regime $\epsilon<1$~\cite{Carlon}. As in
the pure model, the system undergoes a single quantum phase transition with
the control parameter
   \begin{equation}
     \delta=\overline{\ln J}-\overline{\ln h}.
   \end{equation}
SDRG shows that the inter-chain couplings $K_i$ and $g_i$ are irrelevant on the
critical line $\delta=0$, i.e. the random Ashkin-Teller model behaves as two
uncoupled random Ising chains. The critical behavior is therefore governed by
the Fisher infinite-randomness fixed point with the critical exponents
$(\ref{ExposantIRFP})$. However, for finite disorder strength, a strong
cross-over is observed numerically between the pure fixed point and this
infinite-randomness fixed point. The regime $\epsilon>1$ of the random
Ashkin-Teller model was only studied more recently using SDRG~\cite{Vojta1,%
Vojta2}. The phase diagram is qualitatively the same as the pure Ashkin-Teller
model, in particular the two Ising lines still meets at a tricritical point
located at $\delta=0$ and $\epsilon=1$. When this point is approached
by varying $\delta$, the scaling dimensions (\ref{ExposantIRFP}) of the
infinite-randomness Ising fixed point are recovered. However, when approaching
this point along the half-line $\delta=0$ and $\epsilon>1$, the critical
behavior is governed by different exponents:
    \begin{equation}
      \beta={6-2\sqrt 5\over 1+\sqrt 7},\quad
      \nu={8\over 1+\sqrt 7}.
    \end{equation}
Note that the ratio $\beta/\nu$ is unchanged, a property sometimes referred
to as weak universality. Between the two Ising lines in the regime $\epsilon>1$,
SDRG indicates the existence of a double Griffiths phase: magnetization
behaves as in the disordered Griffiths phase of the random Ising chain but
polarization as in the ordered Griffiths phase.
\\

In the rest of the paper, new data of both regimes $\epsilon<1$ and
$\epsilon>1$ obtained by DMRG are presented and discussed. While only
the critical point was considered in~\cite{Carlon}, we are interested in
the out-of-critical region of the phase diagram and especially in the
Griffiths phases when $\epsilon>1$. In the first
section, details about the implementation of the model and the parameters
used for numerical computations are presented. In the second section,
the phase boundaries are determined using integrated autocorrelation times,
and the disorder fluctuations of magnetization and polarization.
They are compared with the behavior of the entanglement entropy
of one half of the lattice with the rest of the system.
In the third section, the spin-spin and polarization-polarization
autocorrelation functions are analyzed more carefully. In particular,
we are interested in the algebraic decay (\ref{AutocorrGrf}) signaling
the existence of a Griffiths phase. Finally, a conclusion follows.

\section{Numerical details}
We have considered a binary distribution of the intra-chain couplings $J_i$:
   \begin{equation}
     \wp(J_i)={1\over 2}\big[\delta(J_i-J_1)+\delta(J_i-J_2)\big]
   \end{equation}
and homogeneous transverse fields $h$ and $g$. Equation (\ref{DefEpsilon})
now reads
    \begin{equation}
      {K_i\over J_i}={g\over h}=\epsilon.
      \label{DefEpsilon2}
    \end{equation}
The critical behavior is expected to be unaffected by this choice. 
Indeed, the probability distributions of $h$ and $g$, initially delta peaks,
will become broader and broader under renormalization so that the
same IRFP will be eventually reached. This choice was made to
minimize the number of disorder configurations.
If $L$ is the lattice size, the number of $J_i$ couplings is $L-1$ with
open boundary conditions so the total number of disorder configurations
is $2^{L-1}$. For small lattice sizes, up to $L=16$, the average over disorder
can be performed exactly and the possibly disastrous consequences of an
under-sampling of rare events can be avoided~\cite{Derrida}.
This strategy is motivated by the fact that we are mainly interested in
Griffiths phases, where the dominant behavior is due to rare disorder
configurations. The drawback is that a precise determination of critical
exponents is more difficult, in contrast to~\cite{Carlon} where the sampling
was limited to 10,000 disorder configurations, allowing for larger lattice
sizes up to $L=32$.
We also made additional calculations for a lattice size $L=20$ but with
an average over only 50,000 disorder configurations, randomly chosen among
the 524,288 ones. As we will see, this under-sampling leads to observable
deviations.
\\

For simplicity, we have moreover restricted ourselves to the case
   \begin{equation}
     J_2=1/J_1\ \Leftrightarrow\ \overline{\ln J_i}=0.
   \end{equation}
and we have chosen a strong disorder by setting $J_1=4$ and $J_2=1/4$.
The quantum control parameter is now
   \begin{equation}
     \delta=-\ln h.
   \end{equation}
\\

The model was studied using the time-dependent Density-Matrix Renormalization
Group algorithm~\cite{White,White2,Schollwoeck}.
A rough estimate of the ground state
is first obtained with the so-called Infinite-Size DMRG algorithm. Because
the couplings are inhomogeneous, the system was grown by adding single spins
to one boundary rather than inserting them between the two blocks. 
After this initial Infinite-Size step, the accuracy of the ground state is
improved by performing four sweeps of the Finite-Size algorithm. Since disorder
fluctuations dominate at the IRFP, quantum fluctuations are expected to be
much weaker than in the pure Ashkin-Teller model. For the latter, the expected
critical exponents were recovered by keeping of the order of $m=192$ states when
truncating the Hilbert space of a left or right block in the DMRG algorithm.
For the random Ashkin-Teller model, we fixed the upper limit of this
parameter to $m=64$. The actual number of states was determined dynamically
by imposing a maximal truncation error: $10^{-5}$ during the initial
Infinite-Size step, $10^{-6}$, $10^{-7}$, $10^{-8}$, and $10^{-9}$ during the
four Finite-Size sweeps. Using these parameters, we were able to make
calculations for a large number of quantum control parameters $\delta$
for lattice sizes up to $L=20$ for $\epsilon>1$.
Unfortunately, the Arpack library, used to determine
the ground-state in the truncated Hilbert space, sometimes failed for some
particular disorder configurations. In these cases, the point, and not
simply this disorder configuration, is discarded. For $\epsilon\le 1$, many
calculations failed for $L=16$. Only 27 values of the control quantum
parameter could be completed for $\epsilon=1$, mostly far from the critical
point. Moreover, when successful,
the computation takes a time which increases very fast for $\epsilon<1$.
Since the two Ising chains are uncoupled at the fixed point, the Hilbert
space becomes closer to a tensor product of the
spaces of two Ising chains. Therefore, the number of states to be kept
during the truncation process of the DMRG algorithm should be of the
order of the square of the number of states necessary for a single Ising
chain. For this reason, the largest lattice size considered
in the regime $\epsilon<1$ is only $L=12$.
\\

Average magnetization and polarization densities
   \begin{equation}
     \overline{\langle m\rangle}
     =\overline{\bra 0\sigma_{L/2}^z\ket 0},\quad
     \overline{\langle p\rangle}
     =\overline{\bra 0\sigma_{L/2}^z\tau_{L/2}^z\ket 0},\quad
   \end{equation}
were measured at the center of the chain. $\ket 0$ denotes the
ground state and the over-line bar stands for
the average over disorder. In order to measure non-vanishing averages,
longitudinal magnetic and electric fields were coupled to the
two boundary spins of the chain with the Hamiltonian
    \begin{equation}
      H_1=B\sigma_1^z+E\sigma_1^z\tau_1^z
      +B\sigma_L^z+E\sigma_L^z\tau_L^z
    \end{equation}
to break the two ${\mathbb Z}_2$ symmetries. The convergence of the
DMRG algorithm is also faster when such boundary fields are imposed.
Spin-spin and polarization-polarization connected autocorrelation functions,
defined as
   \begin{eqnarray}
     \overline{A_\sigma(t)}
     &=&\overline{\bra 0\sigma_{L/2}^z(t)\sigma_{L/2}^z(0)\ket 0}
     -\overline{\langle m\rangle^2},\\
     \overline{A_{\sigma\tau}(t)}&=&\overline{\bra 0\sigma_{L/2}^z(t)\tau_{L/2}^z(t)
       \sigma_{L/2}^z(0)\tau_{L/2}^z(0)\ket 0}
     -\overline{\langle p\rangle^2},\nonumber
     \label{DefAutoCorr}
     \end{eqnarray}
were estimated using a discretized imaginary-time evolution operator:
    \begin{equation}
     \overline{A_\sigma(n\Delta t)}=\overline{\left[{\bra 0\sigma_{L/2}^z
           \big(1-H\Delta t\big)^n\sigma_{L/2}^z\ket 0\over \bra 0
           \big(1-H\Delta t\big)^n\ket 0}\right]}-\overline{\langle m\rangle^2}.
    \end{equation}
We have used the value $\Delta t=10^{-3}$ and computed autocorrelation
functions up to $t=10$.

\section{Phase boundaries}
\begin{figure*}[]
\centering
\psfrag{H}[tc][tc][1][0]{$h$}
\psfrag{<m>}[Bc][Bc][1][1]{$\overline{\langle m\rangle}$}
\psfrag{<p>}[Bc][Bc][1][1]{$\overline{\langle p\rangle}$}
\psfrag{K/J=0.25}[Bc][Bc][1][0]{\small $\epsilon=1/4$}
\psfrag{K/J=0.50}[Bc][Bc][1][0]{\small $\epsilon=1/2$}
\psfrag{K/J=1.00}[Bc][Bc][1][0]{\small $\epsilon=1$}
\psfrag{K/J=2.00}[Bc][Bc][1][0]{\small $\epsilon=2$}
\psfrag{K/J=4.00}[Bc][Bc][1][0]{\small $\epsilon=4$}
\includegraphics[width=8cm]{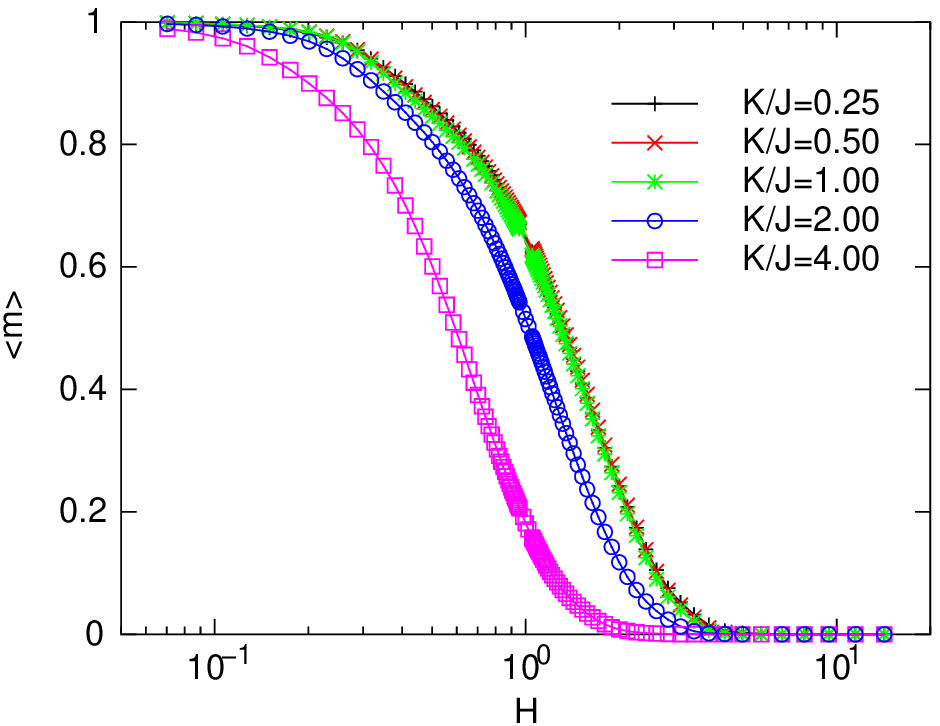}\quad\quad
\includegraphics[width=8cm]{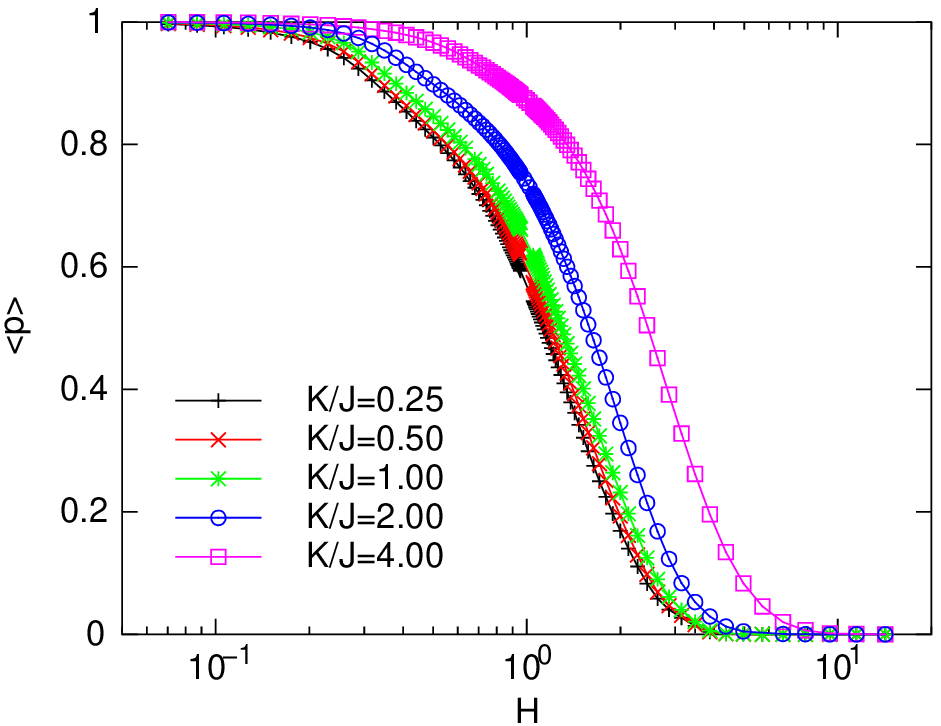}
\caption{Magnetization (left) and polarization (right) of the random
quantum Ashkin-Teller chain versus the transverse field $h$. The different
curves correspond to different values of $\epsilon=K_i/J_i$. The lattice
size is $L=12$.}
\label{fig10b}
\end{figure*}

As discussed in the introduction, the random quantum Ashkin-Teller model
is expected to undergo a single transition when $\epsilon\le 1$ and two
transitions when $\epsilon>1$. This is easily observed on the behavior
of magnetization and polarization, which are the two order parameters
of these two transitions. As seen on figures~\ref{fig10b}, magnetization
and polarization display a fast variation but at different values of
the transverse field $h$, and therefore of the control parameter
$\delta=-\ln h$, when $\epsilon>1$.

However, because of the finite-size of the system, magnetization and
polarization curves are too smooth to provide accurate estimates of the
location of the transitions. Diverging quantities are more convenient for
that purpose and usually preferred in numerical studies. In this section,
we discuss three quantities that diverge, or display a pronounced peak,
at the transitions of the random Ashkin-Teller model.

\subsection{Integrated autocorrelation time}
One of the properties that define criticality is that any characteristic
length or time disappears at a second-order phase transition.
Out-of-criticality, the exponential decay of average spatial correlation
functions $C(r)$ and autocorrelation functions $A(t)$ provides respectively
a correlation length $\xi$ and an autocorrelation time $\xi_t$. In a pure
system, both quantities are expected to diverge as a critical point is
approached. In the random case, a divergence of $\xi$ and $\xi_t$ is expected
in the whole Griffiths phase. However, in a finite system, these divergences
are smoothed and replaced by a finite peak. At large time $t$, connected
autocorrelation functions $\overline{A(t)}$ are dominated by an exponential
decay of the variable $t/\xi_t$. Consequently, their integrals behave as
     \begin{equation}
       \tau=\int_0^{+\infty} \overline{A(t/\xi_t)}dt
       =\xi_t\int_0^{+\infty} \overline{A(u)}du
       \label{DefTau}
      \end{equation}
and, like $\xi_t$, should display a peak.
We have computed the integrated autocorrelation time $\tau$ for spin-spin and
polarization-polarization autocorrelation functions. The upper bound of
the integral (\ref{DefTau}) was replaced by the largest time $t=10$
considered. This approximation has no effect on the
estimate of the autocorrelation time $\tau$ as long as $\xi_t$ is
much smaller than $10$. As will be seen below, this is the
case for the lattice sizes that we considered.
\\

\begin{figure*}
\centering
\psfrag{H}[tc][tc][1][0]{$h$}
\psfrag{taum}[Bc][Bc][1][1]{$\xi_t$}
\psfrag{L=08                }[Br][Bl][1][0]{\small $L=8$}
\psfrag{L=10                }[Br][Bl][1][0]{\small $L=10$}
\psfrag{L=12                }[Br][Bl][1][0]{\small $L=12$}
\psfrag{L=16                }[Br][Bl][1][0]{\small $L=16$}
\psfrag{L=20                }[Br][Bl][1][0]{\small $L=20$}
\psfrag{K/J=0.25}[Bc][Bc][1][0]{$\epsilon=1/4$}
\psfrag{K/J=0.50}[Bc][Bc][1][0]{$\epsilon=1/2$}
\psfrag{K/J=1.00}[Bc][Bc][1][0]{$\epsilon=1$}
\psfrag{K/J=2.00}[Bc][Bc][1][0]{$\epsilon=2$}
\psfrag{K/J=4.00}[Bc][Bc][1][0]{$\epsilon=4$}
\includegraphics[width=20cm]{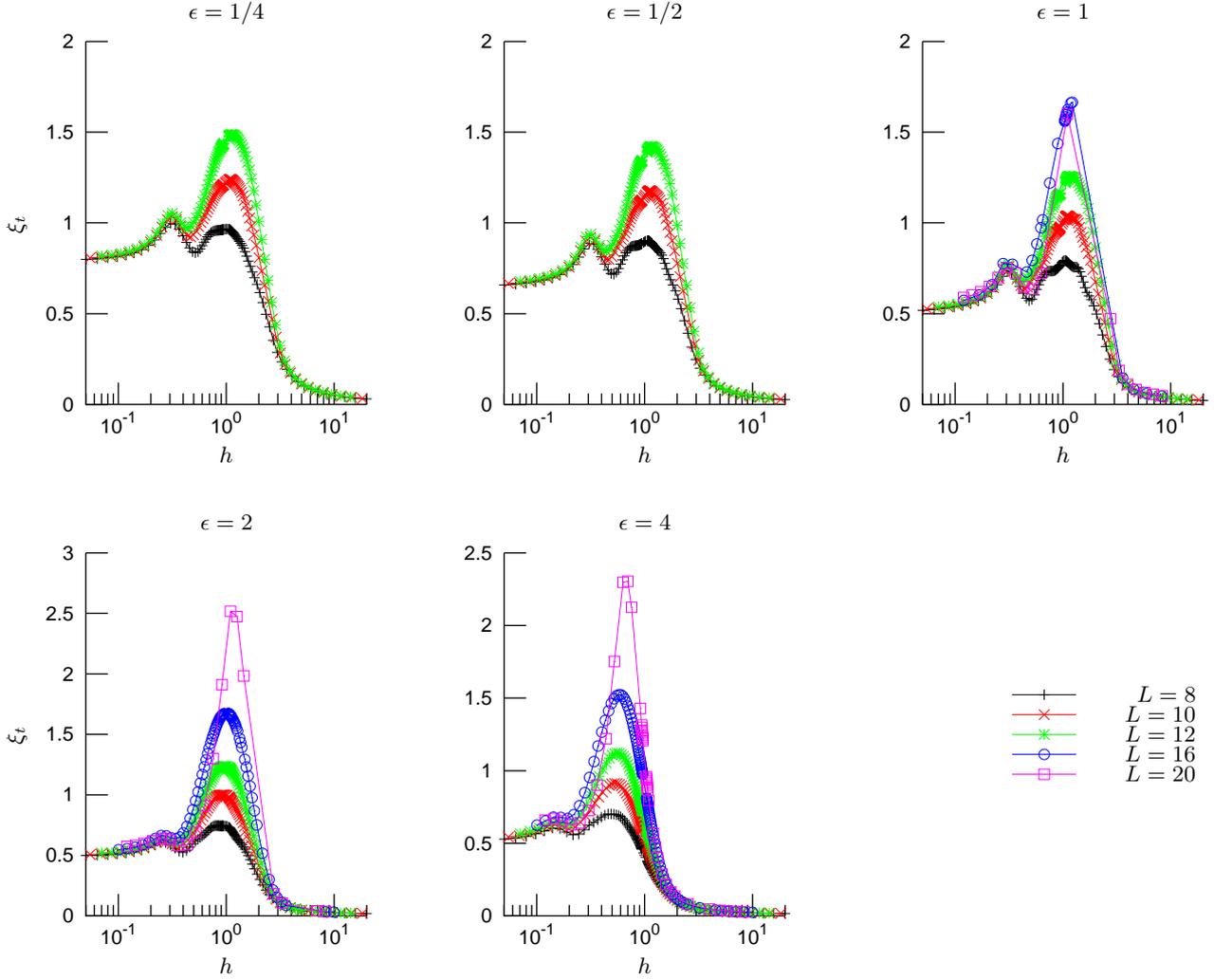}
\caption{Autocorrelation time $\xi_t$ estimated by integration of the average
spin-spin autocorrelation function $A_\sigma(t)$. The different graphs
correspond to different values of $\epsilon$ and the different curves to
different lattice sizes $L$.}
\label{fig12}
\end{figure*}

\begin{figure*}
\centering
\psfrag{H}[tc][tc][1][0]{$h$}
\psfrag{taup}[Bc][Bc][1][1]{$\xi_t$}
\psfrag{L=08                }[Br][Bl][1][0]{\small $L=8$}
\psfrag{L=10                }[Br][Bl][1][0]{\small $L=10$}
\psfrag{L=12                }[Br][Bl][1][0]{\small $L=12$}
\psfrag{L=16                }[Br][Bl][1][0]{\small $L=16$}
\psfrag{L=20                }[Br][Bl][1][0]{\small $L=20$}
\psfrag{K/J=0.25}[Bc][Bc][1][0]{$\epsilon=1/4$}
\psfrag{K/J=0.50}[Bc][Bc][1][0]{$\epsilon=1/2$}
\psfrag{K/J=1.00}[Bc][Bc][1][0]{$\epsilon=1$}
\psfrag{K/J=2.00}[Bc][Bc][1][0]{$\epsilon=2$}
\psfrag{K/J=4.00}[Bc][Bc][1][0]{$\epsilon=4$}
\includegraphics[width=20cm]{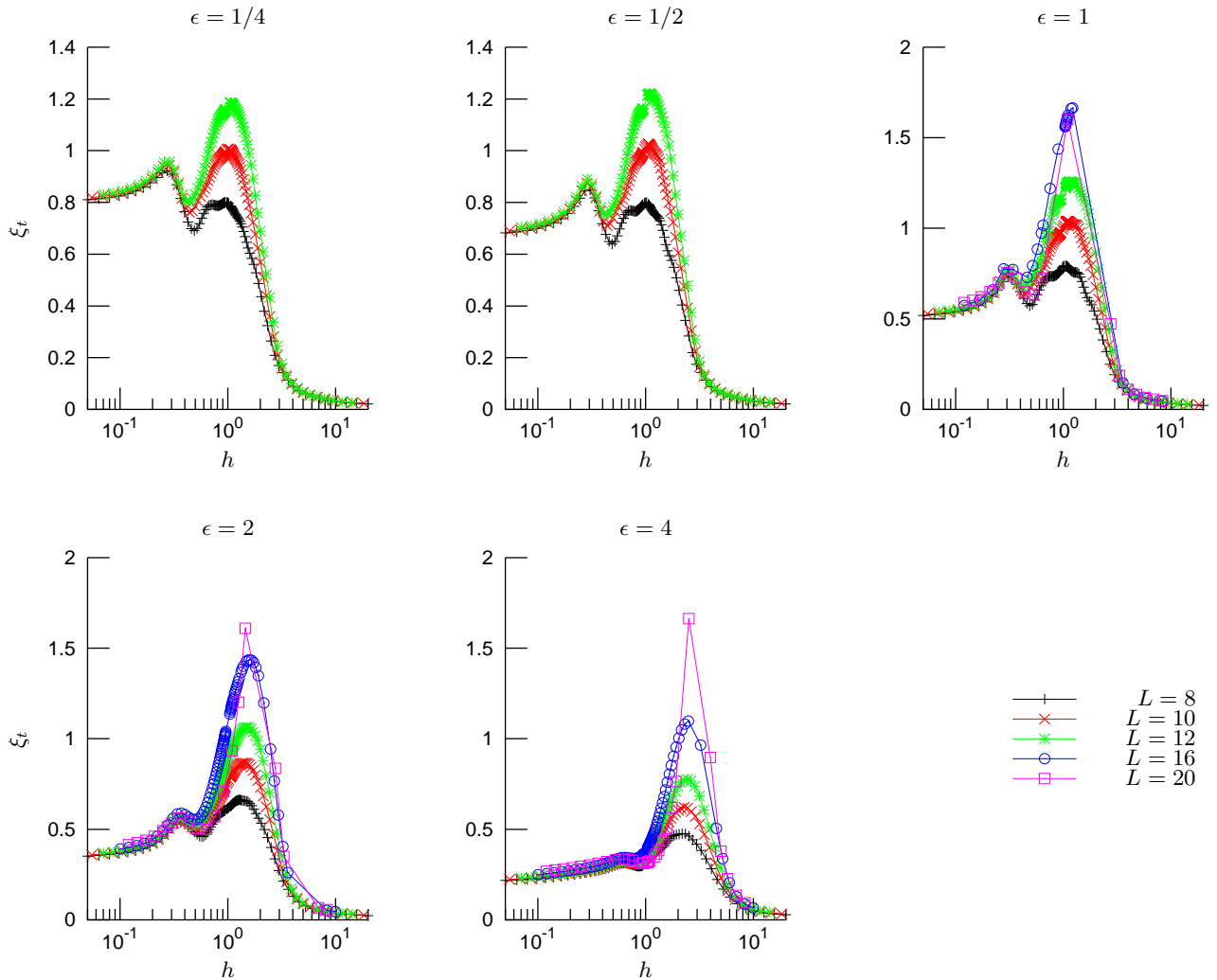}
\caption{Autocorrelation time $\xi_t$ estimated by integration of the
average polarization-polarization autocorrelation function $A_{\sigma\tau}(t)$.
The different graphs correspond to different values of $\epsilon$
and the different curves to different lattice sizes $L$.}
\label{fig13}
\end{figure*}

As can be seen on figures~\ref{fig12} and \ref{fig13}, the integrated
autocorrelation times display two peaks. The first peak occurs at a value
of the transverse field $h$ which is of the same order of magnitude as $J_2$.
Therefore, this peak is probably associated to the ordering transition of the
disorder configurations with a majority of weak couplings $J_2$. However, the
height of this peak does not increase significantly with the lattice size so
one can conjecture that this peak will remain finite in the thermodynamic
limit and is not associated to any phase transition. The height of the
second peak clearly increases with the lattice size. For $\epsilon\le 1$,
the location of the peak is roughly the same for spin-spin and
polarization-polarization autocorrelation times. For $\epsilon>1$, the
data clearly shows that the peak occurs at a positive control parameter
$\delta$, i.e. a transverse field $h<1$, for spin-spin autocorrelation
functions and negative for polarization-polarization ones. This indicates
that the system undergoes an electric phase transition followed, at larger
control parameter, by a magnetic one. This is consistent with the picture
given by magnetization and polarization curves. The location of the two
transitions was predicted by Hrahsheh {\sl et al.} to be $\delta_c=\pm\ln
{\epsilon\over 2}$ for $\epsilon\gg 1$~\cite{Vojta1}.
For $\epsilon=4$, we observe
the two peaks at $\delta_c=-\ln h_c\simeq 0.54$ and $\delta_c\simeq -0.99$
for $L=16$ for instance, still far from $\pm\ln{\epsilon\over 2}
\simeq\pm 0.69$. Moreover, our transitions lines are not symmetric with
respect to the axis $\delta=0$, as required by self-duality. Since
the data was produced by DMRG with a relatively large number of states
and since the averages were made over all disorder configurations, the
deviation can only be the consequence of the relatively small lattice
sizes that could be reached and of the boundary magnetic and electric
fields which favor a Baxter phase and therefore shift the whole phase
diagram.

\begin{figure}
\centering
\psfrag{delta}[Bc][Bc][1][1]{$h$}
\psfrag{K/J}[Bc][Bc][1][1]{$\epsilon$}
\psfrag{L=8}[Br][Br][1][0]{\small $L=8$}
\psfrag{L=10}[Br][Br][1][0]{\small\quad $L=10$}
\psfrag{L=12}[Br][Br][1][0]{\small\quad $L=12$}
\psfrag{L=16}[Br][Br][1][0]{\small\quad $L=16$}
\psfrag{L=20}[Br][Br][1][0]{\small\quad $L=20$}
\includegraphics[width=9cm]{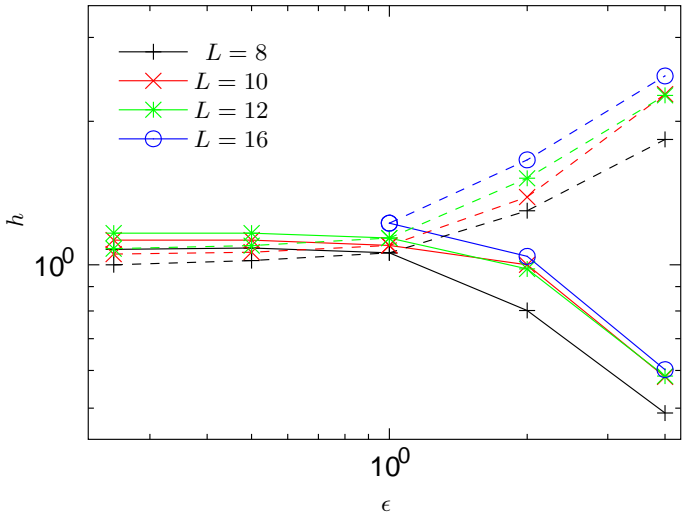}
\caption{Phase diagram in the parameter space $(\epsilon,h)$ obtained
  from spin-spin (continuous lines) and polarization-polarization (dashed lines)
  first moment (\ref{FirstMoment}). The different curves
  correspond to different lattice sizes.}
\label{fig20c}
\end{figure}

We also considered the first moment
     \begin{equation}
       \int_0^{+\infty} t\ \!\overline{A(t)}dt\ \big/\ 
       \int_0^{+\infty} \overline{A(t)}dt
       \label{FirstMoment}
     \end{equation}
that should be equal to the autocorrelation time $\xi_t$ if the connected
autocorrelation function $\overline{A(t)}$ displays a purely exponential decay
$\overline{A(t)}\sim e^{-t/\xi_t}$. Like the autocorrelation time, the first
moment was computed for both spin-spin and polarization-polarization
autocorrelation functions.  When plotted with respect to the transverse fields,
two peaks are observed. Even though the shape of these peaks is not strictly
identical to that of the autocorrelation time (\ref{DefTau}), in particular
the second peak is higher and slightly broader, both quantities behave
in the same way with the transverse field $h$. Therefore, the same
conclusions can be drawn. A reconstructed phase diagram is shown on
figure~\ref{fig20c}. It is qualitatively similar to the one presented
in Ref.~\cite{Vojta1}. However, it is not symmetric under the transformation
$\delta\leftrightarrow -\delta$. As discussed above, finite-size effects
are here strengthened by the boundary magnetic and electric fields that
globally shift the phase diagram.

\subsection{Disorder fluctuations}
\begin{figure*}
\centering
\psfrag{H}[tc][tc][1][0]{$h$}
\psfrag{varm}[Bc][Bc][1][1]{$V_\sigma$}
\psfrag{L=08                }[Br][Bl][1][0]{\small $L=8$}
\psfrag{L=10                }[Br][Bl][1][0]{\small $L=10$}
\psfrag{L=12                }[Br][Bl][1][0]{\small $L=12$}
\psfrag{L=16                }[Br][Bl][1][0]{\small $L=16$}
\psfrag{L=20                }[Br][Bl][1][0]{\small $L=20$}
\psfrag{K/J=0.25}[Bc][Bc][1][0]{$\epsilon=1/4$}
\psfrag{K/J=0.50}[Bc][Bc][1][0]{$\epsilon=1/2$}
\psfrag{K/J=1.00}[Bc][Bc][1][0]{$\epsilon=1$}
\psfrag{K/J=2.00}[Bc][Bc][1][0]{$\epsilon=2$}
\psfrag{K/J=4.00}[Bc][Bc][1][0]{$\epsilon=4$}
\includegraphics[width=20cm]{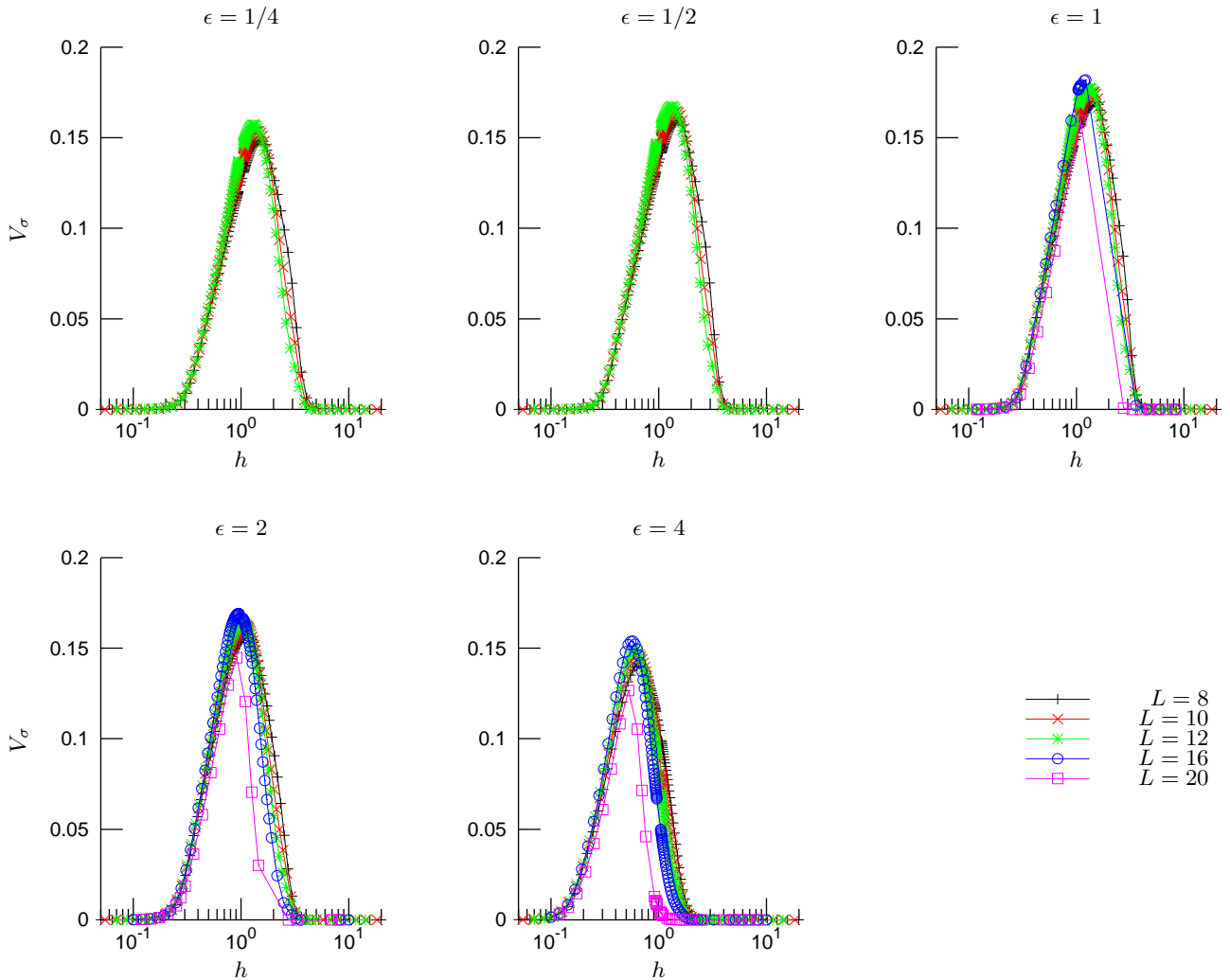}
\caption{Variance of disorder fluctuations of magnetization.
The different graphs correspond to different values of $\epsilon$
and the different curves to different lattice sizes $L$.}
\label{fig18}
\end{figure*}

\begin{figure*}
\centering
\psfrag{H}[tc][tc][1][0]{$h$}
\psfrag{varp}[Bc][Bc][1][1]{$V_{\sigma\tau}$}
\psfrag{L=08                }[Br][Bl][1][0]{\small $L=8$}
\psfrag{L=10                }[Br][Bl][1][0]{\small $L=10$}
\psfrag{L=12                }[Br][Bl][1][0]{\small $L=12$}
\psfrag{L=16                }[Br][Bl][1][0]{\small $L=16$}
\psfrag{L=20                }[Br][Bl][1][0]{\small $L=20$}
\psfrag{K/J=0.25}[Bc][Bc][1][0]{$\epsilon=1/4$}
\psfrag{K/J=0.50}[Bc][Bc][1][0]{$\epsilon=1/2$}
\psfrag{K/J=1.00}[Bc][Bc][1][0]{$\epsilon=1$}
\psfrag{K/J=2.00}[Bc][Bc][1][0]{$\epsilon=2$}
\psfrag{K/J=4.00}[Bc][Bc][1][0]{$\epsilon=4$}
\includegraphics[width=20cm]{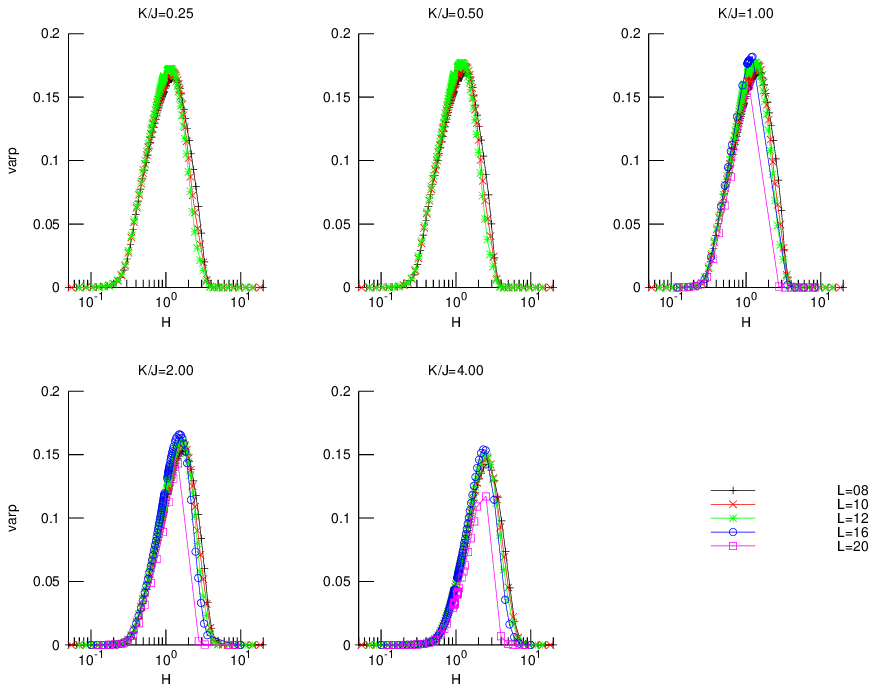}
\caption{Variance of disorder fluctuations of polarization.
The different graphs correspond to different values of $\epsilon$
and the different curves to different lattice sizes $L$.}
\label{fig19}
\end{figure*}

In a random system, any thermodynamic average $\overline{\langle X\rangle}$
is the result of a quantum average
    \begin{equation}
    \langle X\rangle=\bra{\psi_0[J_i,K_i]}X\ket{\psi_0[J_i,K_i]}
    \end{equation}
followed by an average over coupling configurations
    \begin{equation}
      \overline{\langle X\rangle}=\int
      \bra{\psi_0[J_i,K_i]}X\ket{\psi_0[J_i,K_i]}
      \wp(\{J_i,K_i\})\prod_i dJ_idK_i
      \end{equation}
where $\ket{\psi_0[J_i,K_i]}$ is the ground state of the system
for a given coupling configuration $\{J_i,K_i\}$ and $\wp(\{J_i,K_i\})$
the probability of this configuration. At an IRFP,
disorder fluctuations dominate over quantum fluctuations. The strength
of the former can be measured by the variance
    \begin{equation}
     V_X=\overline{\langle X\rangle^2}-\overline{\langle X\rangle}^2.
    \end{equation}
We computed this quantity for both magnetization ($V_\sigma$) and polarization
($V_{\sigma\tau}$). As can be seen on figures~\ref{fig18} and \ref{fig19},
the variances $V_\sigma$ and $V_{\sigma\tau}$ are numerically very
stable. They vanish at high and low transverse fields $h$ and display a
well-defined single peak. In particular, there is no second peak at $h\sim
J_2$. The locations of the maxima of the peaks are accurately determined
and are in agreement with the ones estimated from autocorrelation times.
The same conclusions can be drawn: the magnetic and electric transitions
occur at very close control parameters $\delta$, probably the same, for
$\epsilon\le 1$, while a finite shift is observed for $\epsilon>1$. Even
though only a weak dependence on the lattice size $L$ of $V_\sigma$ and
$V_{\sigma\tau}$ is observed on figures~\ref{fig18} and \ref{fig19}, a
systematic finite-size shift is present. For $\epsilon\le 1$, the distance
between the two critical lines decreases when the lattice size $L$ increases,
in agreement with the prediction of a unique transition.
The coincidence of the maxima of the autocorrelation times with those
of the disorder fluctuations shows that the phase transition is induced
by disorder fluctuations, rather than quantum fluctuations, as expected
at an IRFP.

As can be noticed on figures~\ref{fig18} and~\ref{fig19}, the height
of the peaks of the variance of disorder fluctuations increases slightly
with the lattice size, at least for $L\le 16$. The data for the lattice size
$L=20$ display indeed a smaller peak. This lattice size is the only one for
which the average has not been computed over all possible disorder
configurations but only over a subset ($\sim 10\%$) of them.
The smaller peak for $L=20$ is therefore probably due to an under-sampling
of the dominant configurations at the critical point. $50,000$ is still,
at least for certain quantities, a too small number of disorder configurations.
In the following, data for $L=20$ should be taken with more care than
smaller lattice sizes, for which an exact average over disorder was performed.

\subsection{Entanglement entropy}
When the degrees of freedom of the system can be divided into two subsets
$A$ and $B$, and therefore when the Hilbert space can be decomposed as
a tensor product ${\cal H}={\cal H}_A\otimes{\cal H}_B$, the degree of
entanglement of the two sub-blocks is conveniently measured by the von
Neumann entanglement entropy of $A$ with the rest of the system~\cite{Amico}:
   \def\trace{{\rm Tr}\ \!}
   \begin{equation}
     S_A=-\trace_{{\cal H}_A} \rho_A\log\rho_A
   \end{equation}
where $\rho_A$ is the reduced density matrix
   \begin{equation}
     \rho_A=\trace_{{\cal H}_B} \rho
   \end{equation}
and $\rho$ the density matrix of the full system. In the case of a pure state
$\ket\psi$, the latter is the projector $\rho=\ket\psi\bra\psi$. In the
following, we will consider the subset $A$ made of the $\ell$ spins
at the left of the chain. 

\begin{figure*}
\centering
\psfrag{H}[tc][tc][1][0]{$h$}
\psfrag{S(L/2)}[Bc][Bc][1][1]{$S(L/2)$}
\psfrag{L=08                }[Br][Bl][1][0]{\small $L=8$}
\psfrag{L=10                }[Br][Bl][1][0]{\small $L=10$}
\psfrag{L=12                }[Br][Bl][1][0]{\small $L=12$}
\psfrag{L=16                }[Br][Bl][1][0]{\small $L=16$}
\psfrag{L=20                }[Br][Bl][1][0]{\small $L=20$}
\psfrag{K/J=0.25}[Bc][Bc][1][0]{$\epsilon=1/4$}
\psfrag{K/J=0.50}[Bc][Bc][1][0]{$\epsilon=1/2$}
\psfrag{K/J=1.00}[Bc][Bc][1][0]{$\epsilon=1$}
\psfrag{K/J=2.00}[Bc][Bc][1][0]{$\epsilon=2$}
\psfrag{K/J=4.00}[Bc][Bc][1][0]{$\epsilon=4$}
\includegraphics[width=20cm]{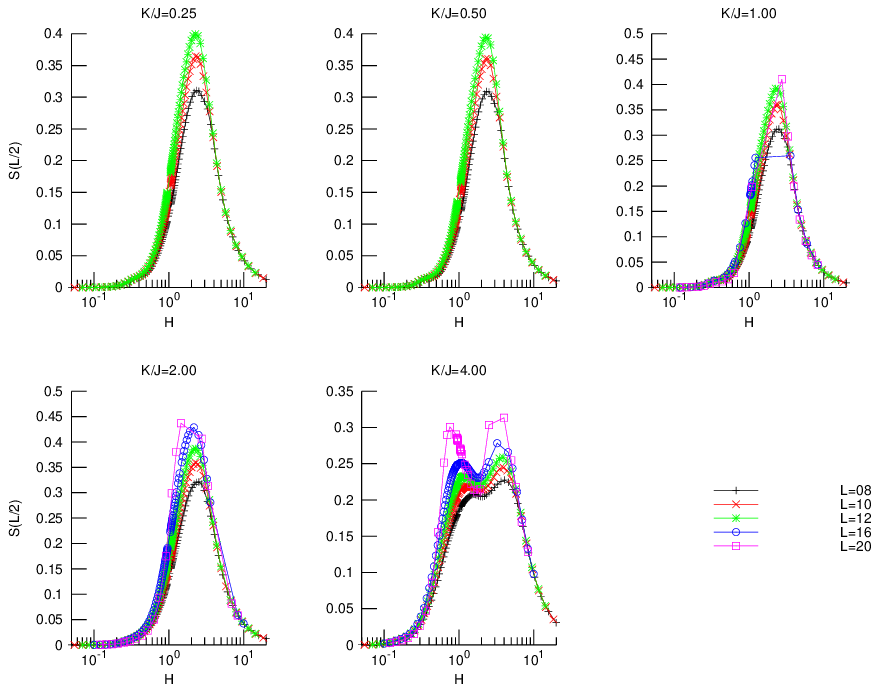}
\caption{Entanglement entropy for an equal partition of the system, i.e.
$\ell=L/2$. The different graphs correspond to different values
of $\epsilon$ and the different curves to different lattice sizes $L$.}
\label{fig21}
\end{figure*}

Entanglement entropy has recently attracted a lot of attention because of
Conformal Field Theory (CFT) predictions at pure critical
points~\cite{Vidal,Cardy}. The predicted logarithmic behavior with $\ell$
is also observed in RTFIM but with a prefactor that involves an effective
central charge $\tilde c={1\over 2}\ln 2$~\cite{Refael}. The entanglement
entropy is also commonly
used in the literature to determine phase boundaries~\cite{Amico}. Indeed,
it is expected to be larger when quantum correlation functions are long-range.
At an IRFP, the entanglement entropy is related to the probability of a
strongly correlated cluster across the boundary between the two
blocks $A$ and $B$. Numerically, the reduced density matrix $\rho_A$ being
computed and diagonalized at each step of the DMRG algorithm, the entanglement
entropy is given without any additional computational effort.
\\

The average entanglement entropy $\overline{S(\ell)}$ of the random quantum
Ashkin-Teller chain is plotted on figure~\ref{fig21} for $\ell=L/2$.
For $\epsilon=4$, two peaks can be observed and interpreted as the signature
of the two phase transitions. As expected, one single peak is present when
$\epsilon\le 1$. However, only one peak can be distinguished in the case
$\epsilon=2$, while autocorrelation time indicates the existence of two
transitions. Because of the finite-size of the system, the expected two peaks
are probably merged into a single one. This scenario is compatible with
what is observed for $\epsilon=4$: what was only a shouldering at the left
of the peak for $L=8$ becomes a second independent peak at $L=20$.
\\

The phase diagram is qualitatively the same as previously constructed.
However, the two peaks are not located at the same position as those
displayed by the integrated autocorrelation times, or the variance of
disorder fluctuations. At $\epsilon=4$, they are instead found at $\delta_c
\sim -0.10$ and $\delta_c\sim -1.33$, far from the estimates $\delta_c
\simeq 0.54$ and $-0.99$. This large difference is probably due to
Finite-Size effects. Indeed, magnetization, polarization and autocorrelation
functions were computed at the center of the lattice, i.e. at the site $L/2$.
In contrast, the entanglement entropy is a global quantity, therefore
more sensitive to the presence of boundary fields.

\begin{figure}
\centering
\psfrag{log L}[tc][tc][1][0]{$\log L$}
\psfrag{S(L/2)}[Bc][Bc][1][1]{$\max_h S(L/2)$}
\psfrag{K/J=0.25}[Bc][Bc][1][0]{$\epsilon=1/4$}
\psfrag{K/J=0.50}[Bc][Bc][1][0]{$\epsilon=1/2$}
\psfrag{K/J=1.00}[Bc][Bc][1][0]{$\epsilon=1$}
\psfrag{K/J=2.00}[Bc][Bc][1][0]{$\epsilon=2$}
\psfrag{K/J=4.00}[Bc][Bc][1][0]{$\epsilon=4$}
\includegraphics[width=9cm]{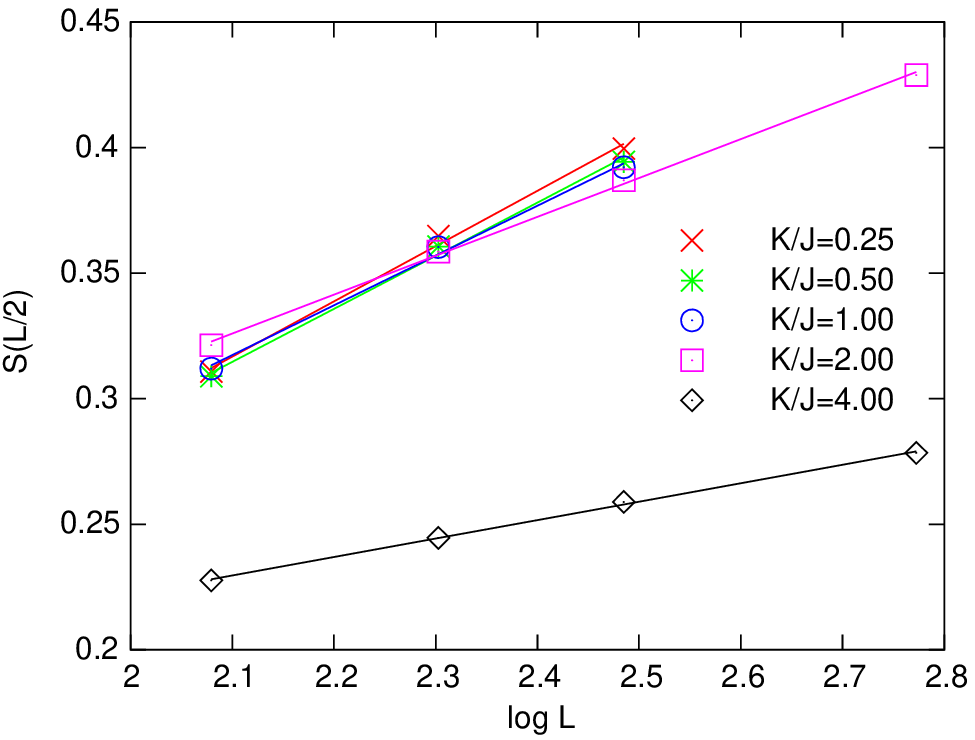}
\caption{Scaling of the maximum of the entanglement entropy $S(L/2)$ for an
  equal partition of the system, i.e. $\ell=L/2$, with the logarithm
  of the lattice size. The different symbols correspond to different values
  of $\epsilon$ and the straight lines are linear fit of the data.}
\label{fig21b}
\end{figure}

CFT predicts that the entanglement entropy of a block of size $\ell$
behaves as~\cite{Cardy}
   \be S(\ell)=\rho c\ln\Big[{L\over\pi a}\sin{\pi\ell\over L}\Big]
   +{\rm Cst.},  \label{EqEntropieCFT}\ee
   where $c$ is the central charge and $\rho$ is equal to $1/3$ for
   periodic boundary conditions and $1/6$ for open
boundaries. However, this relation was obtained on a finite but continuous
manifold and not on a lattice. Therefore, it is only poorly verified
by our numerical data, for which strong lattice effects are still present.
Nevertheless, the predicted dependence on the lattice size $L$ is well
reproduced by the numerical data. For an equal partition of the system,
i.e. when plugging $\ell=L/2$ into (\ref{EqEntropieCFT}),
the entanglement entropy $S(L/2)$ is expected to be
a linear function of $\ln L$ with a slope $\rho c$. The numerical
data at the maximum of $S(L/2)$ is in good agreement with this prediction
as shown on figure~\ref{fig21b}. This confirms the divergence of the
correlation length as the lattice size and therefore, the occurrence of
a phase transition. Because of the magnetic and electric fields
coupled to the boundaries of the system during the numerical
computations, the CFT predictions for the slope of $S(L/2)$ with $\ln L$
do not apply.

\section{Autocorrelation functions}
As discussed in the introduction, the average connected autocorrelation
functions $\overline{A(t)}$ display three different behaviors according
to the values of the parameters $\delta$ and $\epsilon$, i.e. the position
in the phase diagram. On the critical lines, a slow relaxation
(\ref{AutocorrFP}) depending on the logarithm of $t$ is expected.
In the Griffiths phases, rare regions induce an algebraic decay
(\ref{AutocorrGrf}) of the autocorrelation functions, with an exponent $1/z$.
Finally, away from the Griffiths phases, a more usual exponential decay
is recovered.

\begin{figure*}[]
\centering
\psfrag{t}[tc][tc][1][0]{$t$}
\psfrag{Fit non. lin.}[Bc][Bc][1][1]{$\overline{A_\sigma(t)}$}
\psfrag{ H=0.100}[Br][Br][1][0]{\mysmall $h\!=\!0.100$ ($1.00$)}
\psfrag{ H=0.249}[Br][Br][1][0]{\mysmall $h\!=\!0.249$ ($1.00$)}
\psfrag{ H=0.402}[Br][Br][1][0]{\mysmall $h\!=\!0.402$ ($0.66$)}
\psfrag{ H=0.602}[Br][Br][1][0]{\mysmall $h\!=\!0.602$ ($0.54$)}
\psfrag{ H=0.917}[Br][Br][1][0]{\mysmall $h\!=\!0.917$ ($0.67$)}
\psfrag{ H=1.059}[Br][Br][1][0]{\mysmall $h\!=\!1.059$ ($0.82$)}
\psfrag{ H=1.203}[Br][Br][1][0]{\mysmall $h\!=\!1.203$ ($1.00$)}
\psfrag{ H=1.483}[Br][Br][1][0]{\mysmall $h\!=\!1.483$ ($1.00$)}
\psfrag{ H=2.032}[Br][Br][1][0]{\mysmall $h\!=\!2.032$ (exp.)}
\psfrag{ H=3.235}[Br][Br][1][0]{\mysmall $h\!=\!3.235$ (exp.)}
\psfrag{ H=5.969}[Br][Br][1][0]{\mysmall $h\!=\!5.969$ (exp.)}
\psfrag{ H=9.960}[Br][Br][1][0]{\mysmall $h\!=\!9.960$ (exp.)}
\includegraphics[width=8.85cm]{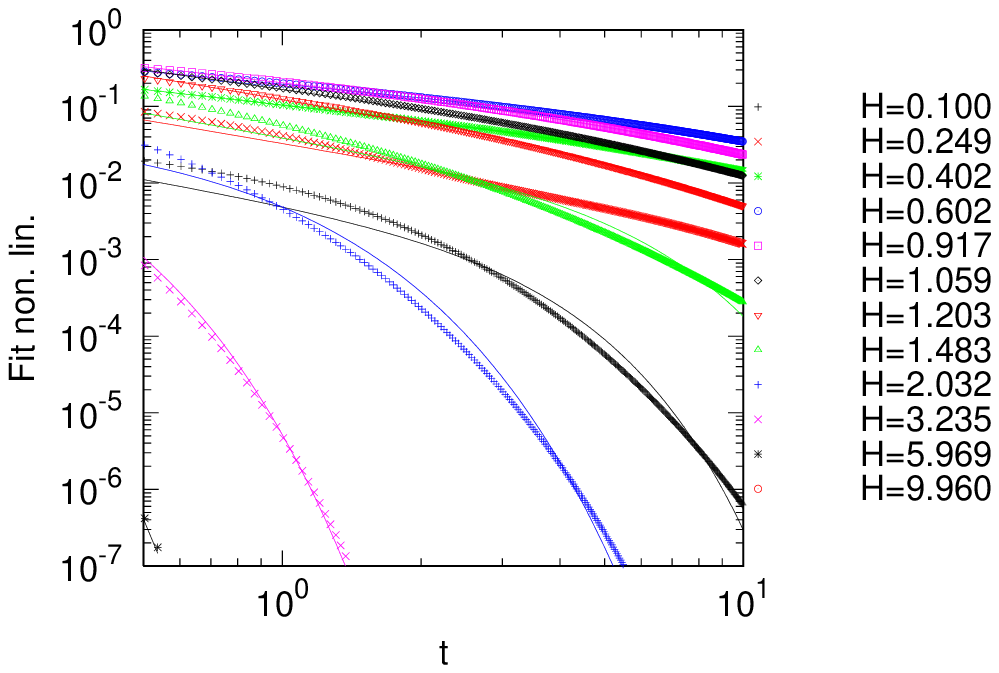}
\psfrag{Fit non. lin.}[Bc][Bc][1][1]{$\overline{A_{\sigma\tau}(t)}$}
\psfrag{ H=0.100}[Br][Br][1][0]{\mysmall $h\!=\!0.100$ (exp.)}
\psfrag{ H=0.249}[Br][Br][1][0]{\mysmall $h\!=\!0.249$ (exp.)}
\psfrag{ H=0.402}[Br][Br][1][0]{\mysmall $h\!=\!0.402$ (exp.)}
\psfrag{ H=0.602}[Br][Br][1][0]{\mysmall $h\!=\!0.602$ ($1.00$)}
\psfrag{ H=0.917}[Br][Br][1][0]{\mysmall $h\!=\!0.917$ ($1.00$)}
\psfrag{ H=1.059}[Br][Br][1][0]{\mysmall $h\!=\!1.059$ ($1.00$)}
\psfrag{ H=1.203}[Br][Br][1][0]{\mysmall $h\!=\!1.203$ ($0.91$)}
\psfrag{ H=1.483}[Br][Br][1][0]{\mysmall $h\!=\!1.483$ ($0.82$)}
\psfrag{ H=2.032}[Br][Br][1][0]{\mysmall $h\!=\!2.032$ ($0.63$)}
\psfrag{ H=3.235}[Br][Br][1][0]{\mysmall $h\!=\!3.235$ ($0.64$)}
\psfrag{ H=5.969}[Br][Br][1][0]{\mysmall $h\!=\!5.969$ ($1.00$)}
\includegraphics[width=8.85cm]{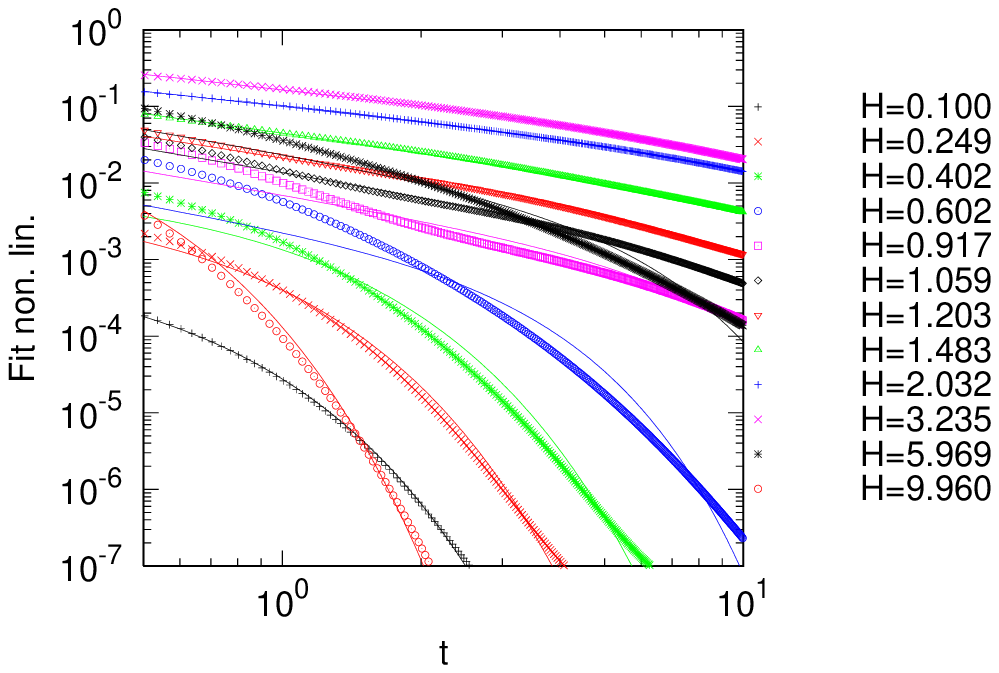}
\caption{Spin-spin (left) and polarization-polarization (right)
autocorrelation functions of the random quantum Ashkin-Teller chain versus
time $t$. The different graphs correspond to different
values of the transverse field $h$. The continuous lines correspond
to a fit, either with the {\sl ansatz} (\ref{AnsatzA}) or with an exponential.
In the legend, the number between parenthesis after the transverse field $h$
is the estimate of $1/z$ given by the fit. When the best fit is the
exponential, {\sl exp} is indicated instead.}
\label{fig16-7}
\end{figure*}

The algebraic decay of autocorrelation functions in the Griffiths phases
has been observed numerically in the case of the random quantum Ising chain
by exploiting the mapping onto a gas of free fermions~\cite{Young}.
The lattice sizes that we were able to reach with DMRG being much smaller,
such an algebraic decay of the spin-spin or polarization-polarization
autocorrelation functions could not be observed for the random
Ashkin-Teller model. A purely algebraic behavior is indeed expected
to hold only in the large-time limit $t\gg 1$ and in the thermodynamic limit
$L\gg 1$. A transient regime may be observed for small times $t$ while,
for large times $t$, the finite-size of the system may induce an exponential
decay of autocorrelation functions. Usually, one looks for an intermediate
regime in the numerical data where the asymptotic behavior holds. No such
intermediate regime could be found for both the spin-spin or
polarization-polarization autocorrelation functions. This is particularly
clear when plotting an effective exponent ${d\ln \overline{A(t)}\over d\ln t}$
versus $t$. For values of the transverse field $h$ expected to be in the
Griffiths
phases, two non-algebraic regimes, where the effective exponent varies with
$t$, are observed at short and large times. But in between, no plateau
corresponding to a purely algebraic decay could be distinguished.
\\

To fit our numerical data, we used an extended expression of the one
proposed by Rieger {\sl et al.} for autocorrelation functions in a Griffiths
phase~\cite{RiegerYoung}. The assumptions are the same: in the paramagnetic
phase, the probability of an ordered region of linear size $\ell$ scales as
$\wp(\ell)\sim e^{-c\ell}$ and its tunneling time is $\tau(\ell)\sim
e^{\sigma'\ell}$. In a finite system of width $L$, the linear size of rare
regions is bounded by $L$ so the average autocorrelation function reads
   \begin{eqnarray}
     \overline{A(t)}
     =\overline{e^{-t/\tau}}
     &=&\int_0^L \wp(\ell)e^{-t/\tau(\ell)}d\ell \nonumber\\
     &=&{t^{-c/\sigma'}\over\sigma'}\int_{te^{-\sigma'L}}^t v^{c/\sigma'-1}e^{-v}dv
     \nonumber\\
     &=&{t^{-1/z}\over\sigma'}\big[\gamma(1/z,t)
       -\gamma(1/z,te^{-\sigma'L})\big]
   \end{eqnarray}
where $v=te^{-\sigma'\ell}$, $\sigma'/c=z$ is the dynamical exponent,
and $\gamma(a,x)$ is the incomplete gamma function.
In the limit of large time $t$ and lattice size $L$, one recovers the
prediction $\overline{A(t)}={\Gamma(1/z)\over\sigma'}t^{-1/z}$ obtained
in the saddle-point approximation.
\\ 

The numerical estimates of the connected autocorrelation functions were
fitted with the 4-parameter non-linear {\sl ansatz}
\begin{equation}
  \overline{A(t)}=a_1t^{-a_2}|\gamma(a_2,a_3t)-\gamma(a_2,a_4t)|.
  \label{AnsatzA}
  \end{equation}
The bounds $0<a_2\le 1$ were imposed during the fitting procedure. The quality
of the fit was quantified using the mean-square deviation $\chi^2$. Because
the boundaries of the Griffiths phases are not known with a good accuracy,
the data were also fitted with an exponential $\overline{A(t)}=a_1e^{-a_2t}$.
Spin-spin and polarization-polarization autocorrelation functions are
plotted respectively on figures \ref{fig16-7} for various transverse fields
$h$ at $\epsilon=4$. The continuous lines correspond to the best fit,
Eq. (\ref{AnsatzA}) or exponential, i.e. the one with the smallest mean square
deviation $\chi^2$. The inverse $1/z$ of the dynamical exponent is indicated
in the legend when (\ref{AnsatzA}) is the best fit, while {\sl exp} indicates
a fit with an exponential. As can be seen on the figures, the data are nicely
reproduced by an exponential decay for large and small transverse fields $h$.
Close to the transition point,
the best fit is obtained with (\ref{AnsatzA}), which means
that the corresponding range of transverse fields is in a Griffiths phase.
As expected, when $\epsilon\le 1$, these phases are centered around $h=1$
and their boundaries are similar for spin-spin and polarization-polarization
autocorrelation functions. For $\epsilon=4$, the Griffiths phase is shifted
to smaller values of the transverse field for spin-spin autocorrelation
functions and to larger ones for polarization-polarization autocorrelations.
For $\epsilon=2$, the shift is seen only for the polarization-polarization
autocorrelation functions. It was also the case for the peak of the
autocorrelation time (figure~\ref{fig12}).
At the boundaries of the Griffiths phases, the data is not well fitted,
neither by an exponential form nor by the {\sl ansatz} (\ref{AnsatzA}).
When the best fit is obtained with the {\sl ansatz} (\ref{AnsatzA}), the
dynamical exponent takes a value $z=1$, i.e. saturating the imposed bound
$z\le 1$. The deviation between the fit and the numerical data is clearly
visible on figures \ref{fig16-7}. The transverse fields for which such a
deviation occurs, are probably in a region of cross-over where the
autocorrelation functions display a more complex behavior.

\begin{figure*}
\centering
\psfrag{H}[tc][tc][1][0]{$h$}
\psfrag{1/zm}[Bc][Bc][1][1]{$1/z$}
\psfrag{L=08                }[Br][Bl][1][0]{\small $L=8$}
\psfrag{L=10                }[Br][Bl][1][0]{\small $L=10$}
\psfrag{L=12                }[Br][Bl][1][0]{\small $L=12$}
\psfrag{L=16                }[Br][Bl][1][0]{\small $L=16$}
\psfrag{L=20                }[Br][Bl][1][0]{\small $L=20$}
\psfrag{K/J=0.25}[Bc][Bc][1][0]{$\epsilon=1/4$}
\psfrag{K/J=0.50}[Bc][Bc][1][0]{$\epsilon=1/2$}
\psfrag{K/J=1.00}[Bc][Bc][1][0]{$\epsilon=1$}
\psfrag{K/J=2.00}[Bc][Bc][1][0]{$\epsilon=2$}
\psfrag{K/J=4.00}[Bc][Bc][1][0]{$\epsilon=4$}
\includegraphics[width=20cm]{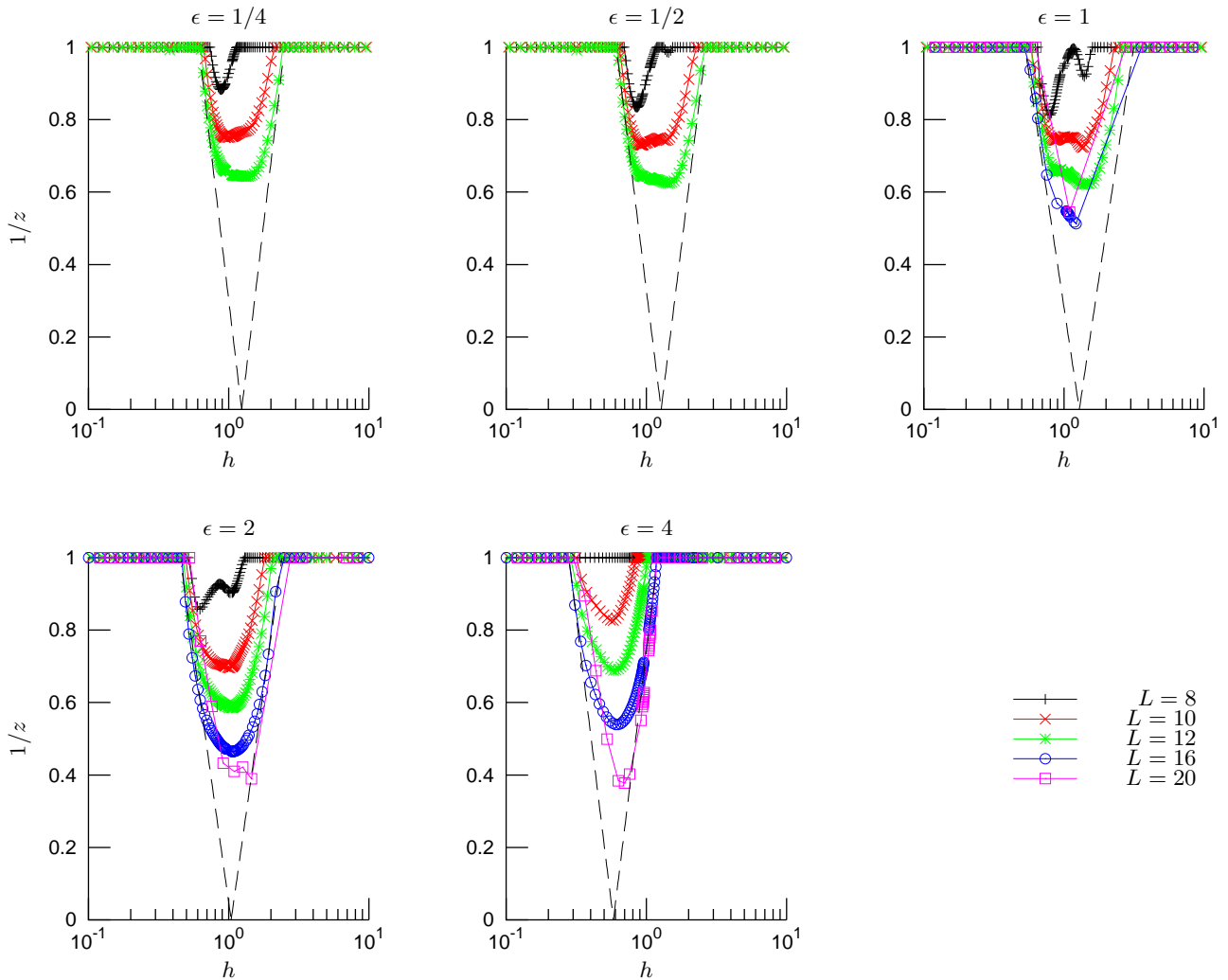}
\caption{Inverse of the dynamical exponent $z$ estimated by a fit
  of the spin-spin autocorrelation functions with Eq. (\ref{AnsatzA})
  and plotted versus the transverse field $h$. The different curves correspond
  to different lattice sizes and the different graphs to different values
  of $\epsilon$.}
\label{fig16c}
\end{figure*}

\begin{figure*}
\centering
\psfrag{H}[tc][tc][1][0]{$h$}
\psfrag{1/zp}[Bc][Bc][1][1]{$1/z$}
\psfrag{L=08                }[Br][Bl][1][0]{\small $L=8$}
\psfrag{L=10                }[Br][Bl][1][0]{\small $L=10$}
\psfrag{L=12                }[Br][Bl][1][0]{\small $L=12$}
\psfrag{L=16                }[Br][Bl][1][0]{\small $L=16$}
\psfrag{L=20                }[Br][Bl][1][0]{\small $L=20$}
\psfrag{K/J=0.25}[Bc][Bc][1][0]{$\epsilon=1/4$}
\psfrag{K/J=0.50}[Bc][Bc][1][0]{$\epsilon=1/2$}
\psfrag{K/J=1.00}[Bc][Bc][1][0]{$\epsilon=1$}
\psfrag{K/J=2.00}[Bc][Bc][1][0]{$\epsilon=2$}
\psfrag{K/J=4.00}[Bc][Bc][1][0]{$\epsilon=4$}
\includegraphics[width=20cm]{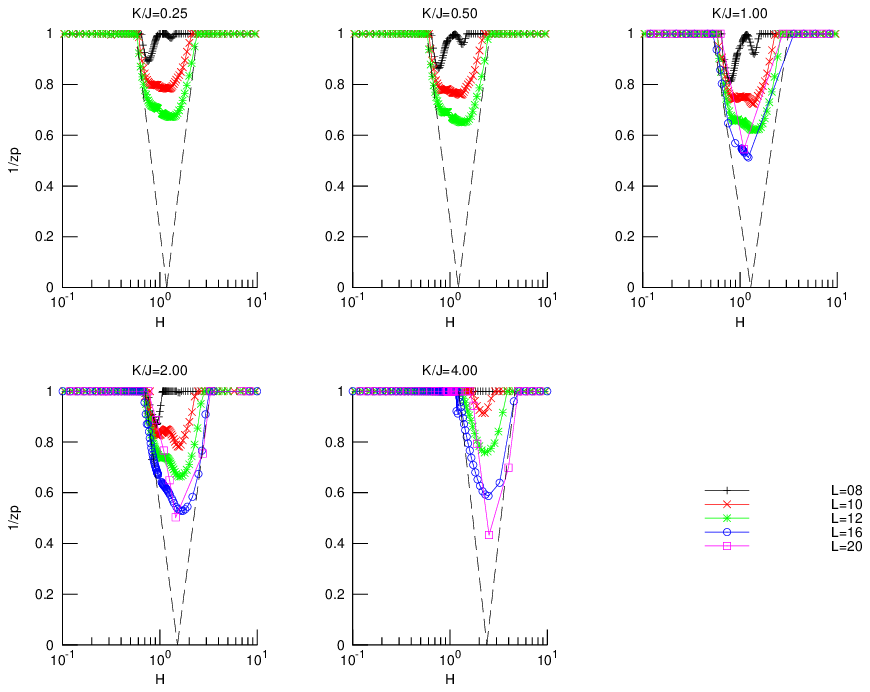}
\caption{Inverse of the dynamical exponent $z$ estimated by a fit of the
polarization-polarization autocorrelation function with Eq.~(\ref{AnsatzA})
and plotted versus $h$. The different curves correspond to different
lattice sizes and the different graphs to different values of $\epsilon$.}
\label{fig17c}
\end{figure*}

On figures~\ref{fig16c} and \ref{fig17c}, the inverse of the dynamical
exponents $z$ is plotted versus the transverse field $h$. As conjectured
in Ref.~\cite{Vojta1}, the dynamical exponents display a peak centered at
the corresponding critical point, i.e. at the magnetic transition for
the dynamical exponent of spin-spin autocorrelation functions and
electric transition for polarization-polarization autocorrelations.
As already observed for other peaked quantities, the two transitions
occur at the same control parameters for $\epsilon\le 1$ and are
separated for $\epsilon>1$. Note that the maxima of the dynamical
exponents are found at the locations of those of the autocorrelation
times and of the variance of disorder fluctuations. Between these
two transitions lines, there is therefore a double Griffiths phase,
i.e. a disordered Griffiths phase in the magnetic sector and an ordered
one in the electric sector where both dynamical exponents $z$ are
larger than 1. However, as seen on figures~\ref{fig16c} and \ref{fig17c},
these Griffiths phase are not infinite but have a finite extension,
because of the binary distribution of the couplings $J_i$
and $K_i$. For $\epsilon=4$, it is observed that the magnetic and electric
Griffiths phases still overlap. Nevertheless, it will probably not
be the case anymore for larger values of $\epsilon$.
\\

In the random Ising chain, the dynamical exponent was shown to
behave as $z\sim 1/2|\delta|$ in the Griffiths phase~\cite{Igloi2001}.
A similar behavior
seems to be also reasonable in the case of the random Ashkin-Teller
chain, as can be seen on figures~\ref{fig16c} and \ref{fig17c}.
The boundaries $\delta_+=-\ln h_+$ and $\delta_-=-\ln h_-$ of the Griffiths
phase were first estimated respectively as the first and last points
with a dynamical exponent $z>1$. The critical point is assumed to
be located at $\delta_c=(\delta_++\delta_-)/2$. The two dashed
lines plotted on figures~\ref{fig16c} and \ref{fig17c} simply correspond
to straight lines $1/z(\delta)=(\delta-\delta_c)/(\delta_+-\delta_c)$
for $\delta\in [\delta_c;\delta_+]$ and $1/z(\delta)=(\delta-\delta_c)
/(\delta_--\delta_c)$ for $\delta\in [\delta_-;\delta_c]$. The slope
is not equal to two, as in the Ising model, but is in the range $1-1.5$.
As the lattice size is increased, the numerical data seem to accumulate
on these straight lines, at least at the boundaries of the Griffiths
phase. In the neighborhood of the critical point, much larger lattice
sizes would be necessary to test this linear behavior of $1/z$.
In the case of $L=20$, the dynamical exponent seems to be over-estimated
for $h<h_c$. Again, this may be explained by the under-sampling
already observed with disorder fluctuations of magnetization and
polarization.

\section{Conclusions}
The random quantum Ashkin-Teller chain has been studied by means of
time-dependent Density Matrix Renormalization Group. The average
over all possible disorder configurations was performed for $L\le 16$.
For $L=20$, a partial average is observed to induce an under-sampling
of disorder fluctuations of magnetization and polarization. Such
partial averages are commonly used in the literature in the study
of random systems. Our data show that they should be considered with
great care, especially in the quantum case.
\\

The analysis of integrated autocorrelation times and of the variance of
disorder fluctuations leads to a phase diagram qualitatively in
agreement with the one conjectured by Hrahsheh {\sl et al.} on the basis
of SDRG~\cite{Vojta1}. However, finite-size effects are large,
especially for entanglement entropy, and our lattice sizes are too small
to allow for an accurate extrapolation in the thermodynamic limit.
The coincidence of the maxima of disorder fluctuations with the critical
lines confirms that the phase transition is governed by disorder
fluctuations, and not by quantum fluctuations. Nevertheless, the
divergence of the entanglement entropy as the logarithm of the
lattice size is recovered, as in pure quantum chains.
In the regime $\epsilon>1$, the existence of a double Griffiths phase
is confirmed. Using an original method to take into account finite-size
effects, the two dynamical exponents, associated to the algebraic
decay of spin-spin and polarization-polarization autocorrelation
functions respectively, could be computed. They display the expected behavior
in a Griffiths phase: a peak centered at the magnetic or electric transition.
Furthemore, it seems reasonnable to assume that they diverge in the
thermodynamic limit as $z(\delta)\sim 1/|\delta|$.


\begin{acknowledgement}
  It is our pleasure to gratefully thank C\'ecile Monthus for
  discussions and for having pointing out some useful references
  on the topic.
\end{acknowledgement}

\end{document}